\documentclass{nature}
\usepackage[ngerman,british]{babel}
\usepackage{graphicx}
\usepackage{amsmath}
\usepackage{epstopdf}
\usepackage{color}
\usepackage[dvipsnames]{xcolor}
\usepackage{url}
\usepackage[normalem]{ulem}
\usepackage{times}
\usepackage[mathlines]{lineno}
\usepackage{bm}
\usepackage{MnSymbol}
\usepackage{setspace}
\usepackage{caption}
\usepackage{tikz}
\usetikzlibrary{shapes}
\usepackage{etoolbox} 
\usepackage[autostyle]{csquotes}

\useshorthands{"}
\addto\extrasbritish{\languageshorthands{ngerman}}

\newcommand{\markerone}{\raisebox{0.5pt}{\tikz{\node[draw,scale=0.4,circle,fill=black!100!black](){};}}}
\newcommand{\markertwo}{\raisebox{0.5pt}{\tikz{\node[draw,scale=0.4,regular polygon, regular polygon sides=4,fill=black!90!black](){};}}}
\newcommand{\markertwob}{\raisebox{0.5pt}{\tikz{\node[draw,scale=0.4,regular polygon, regular polygon sides=4,fill=black!70!,black!70!](){};}}}
\newcommand{\markerthree}{\raisebox{0pt}{\tikz{\node[draw,scale=0.3,regular polygon, regular polygon sides=3,fill=black!80!,rotate=180](){};}}}
\newcommand{\markerthreeb}{\raisebox{0pt}{\tikz{\node[draw,scale=0.3,regular polygon, regular polygon sides=3,fill=black!45!,black!45!,rotate=180](){};}}}
\newcommand{\markerthreec}{\raisebox{0pt}{\tikz{\node[draw,scale=0.3,regular polygon, regular polygon sides=3,fill=black!70!,black!70!,rotate=180](){};}}}
\newcommand{\markerfive}{\raisebox{0pt}{\tikz{\node[draw,scale=0.3,regular polygon, regular polygon sides=3,fill=black!55,black!55,rotate=90](){};}}}
\newcommand{\markersix}{\raisebox{0pt}{\tikz{\node[draw,scale=0.3,regular polygon, regular polygon sides=3, fill=black!35,black!35,rotate=270](){};}}}
\newcommand{\markersixb}{\raisebox{0pt}{\tikz{\node[draw,scale=0.3,regular polygon, regular polygon sides=3, fill=black!55,black!55,rotate=270](){};}}}

\newcommand{\ud}{\,\mathrm{d}}

\newcommand*\linenomathpatchAMS[1]{%
	\expandafter\pretocmd\csname #1\endcsname {\linenomathAMS}{}{}%
	\expandafter\pretocmd\csname #1*\endcsname{\linenomathAMS}{}{}%
	\expandafter\apptocmd\csname end#1\endcsname {\endlinenomath}{}{}%
	\expandafter\apptocmd\csname end#1*\endcsname{\endlinenomath}{}{}%
}
\expandafter\ifx\linenomath\linenomathWithnumbers
\let\linenomathAMS\linenomathWithnumbers
\patchcmd\linenomathAMS{\advance\postdisplaypenalty\linenopenalty}{}{}{}
\else
\let\linenomathAMS\linenomathNonumbers
\fi

\linenomathpatchAMS{gather}
\linenomathpatchAMS{multline}
\linenomathpatchAMS{align}
\linenomathpatchAMS{alignat}
\linenomathpatchAMS{flalign}

\title{The hidden universality of movement in cities}

\author{Markus Schl\"apfer$^{1,2,3}$, Michael Szell$^{2,4,5}$, Hadrien Salat$^{3,6}$, Carlo Ratti$^{2}$ \& Geoffrey B. West$^{1,*}$}

\begin{document} 

\maketitle 

\vspace{-0.5cm}
\begin{affiliations}
\item {\small Santa Fe Institute, Santa Fe, New Mexico 87501, USA}
\item {\small Senseable City Laboratory, Massachusetts Institute of Technology, Cambridge, Massachusetts 02139, USA}
\item {\small Future Cities Laboratory, Singapore-ETH Centre, ETH Zurich, Singapore 138602, Singapore}
\item {\small IT University of Copenhagen, 2300 Copenhagen, Denmark}
\item {\small ISI Foundation, 10126 Turin, Italy}
\item {\small Sociology and Economics of Networks and Services, Orange Labs, 92320 Ch\^{a}tillon, France}
\vspace{0.25cm}
\item[$^*$]{\small \normalfont{Corresponding author (gbw@santafe.edu)}\vspace{0.75cm}}
\end{affiliations}


\setlength{\parskip}{1.2em}


\begin{abstract}
The interaction of all mobile species with their environment hinges on their movement patterns: the places they visit and how frequently they go there\cite{pyke1978optimal,jackson2004trail,strandburg2015shared}. In human society, where the prevalent form of cohabitation is in cities, the highly dynamic and diverse movement of people is fundamental to almost every aspect of socio-economic life, including social in"-ter"-ac"-tions\cite{mumford1961cih,wellman1999ngv} or disease spreading\cite{anderson1991,eubank2004disease}, and ultimately is key to the evolution of urban infrastructure\cite{batty2013nsc}, productivity\cite{sveikauskas1975poc}, innovation\cite{rogers2003doi,fujita1999sec,bettencourt2013osc} and technology\cite{glaeser1992gc}. However, despite the crucial role of the spatio-temporal structure of movement in cities, the laws that govern the variation of population flows to specific locations have remained elusive. Here we show that behind the apparent complexity of movement a surprisingly simple universal scaling relation drives the flow of individuals to \textbf{\textit{any}} specific location based on both frequency of visitation and distance travelled. We derive a first principles argument stating that the \mbox{number of visiting} individuals should decrease as an inverse square of the product of visitation frequency and travel distance; or, equivalently, as a power law with exponent \mbox{\boldmath$\approx \! -2$}. Using large-scale data analyses, we demonstrate that population flows obey this theoretical prediction in virtually all tested areas across the globe, ranging from Europe and America to Asia and Africa, regardless of the detailed geographies, cultures or levels of development. The revealed regularity offers unprecedented possibilities for the modelling of mobility fluxes at high spatial and temporal resolution, and it places an important constraint on any theory of movement, spatial organisation and social interaction in cities.
\end{abstract}

Places attract humans for a multitude of individual reasons that range from daily work and social interactions to entertainment and recreation. Consequently, collective visitation patterns typically span a wide range of both temporal and spatial scales, from highly frequent visitors within a nearby neighbourhood to those that travel over several thousand kilometres to visit a specific location once in a lifetime. Indeed, the attractive forces that constantly act on populations drive the volatility of human activities in space and time\cite{roth2010cpc,deville2014dynamic}, giving rise to heterogeneous yet predictable trajectories of individuals\cite{gonzalez2008uih,song2010msp} as well as to remarkable regularities in the aggregate organisation of human agglomerations\cite{zipf1949ple,makse1995mug, batty2013nsc}. Understanding the highly dynamic population flows to different locations is therefore key to elucidating the social mixing and interactions among people\cite{wei2013ddtf,schlapfer2012shi} that catalyse the essential socio-economic activity of a city\cite{bettencourt2007gis}. But until now, social interaction models\cite{lill1891rsa,zipf1946phi,erlander1990gmt,noulas2012tmc,simini2012umm} have confined  their focus to purely spatial aspects of collective population flows, ignoring the importance of their inherent frequency spectrum. Hence, an essential component of urban dynamics critical for understanding human agglomerations has been neglected.

In this paper we address this omission and develop a general theory that predicts that the spatio-temporal structure of movement in cities follows remarkably simple universal rules that are well supported by data from across the globe. Our quantitative model is based on general principles and provides a comprehensive analysis of how often and from how far away people are attracted to any specific location. The complexity of the spatio-temporal movement is reduced to a single dimension, which allows us to decompose the total flow of individuals to any given location into the underlying distribution of both visitation frequencies and travel distances. {Importantly, these predictions are consistent with, and made manifest in a micro-level mechanistic model of individual behaviour constrained by the spatial organization of cities as interacting attraction centers\cite{dong2020spectral}.} To provide real-world evidence for the universality of our predictions, we compiled a large-scale data set that contains detailed trajectories of millions of individuals in highly diverse urban regions on four different continents -- the Greater Boston area in the USA, Portugal, Senegal and Singapore (Supplementary Information section~III).

\begin{figure}
\centering
\includegraphics[width=1\textwidth]{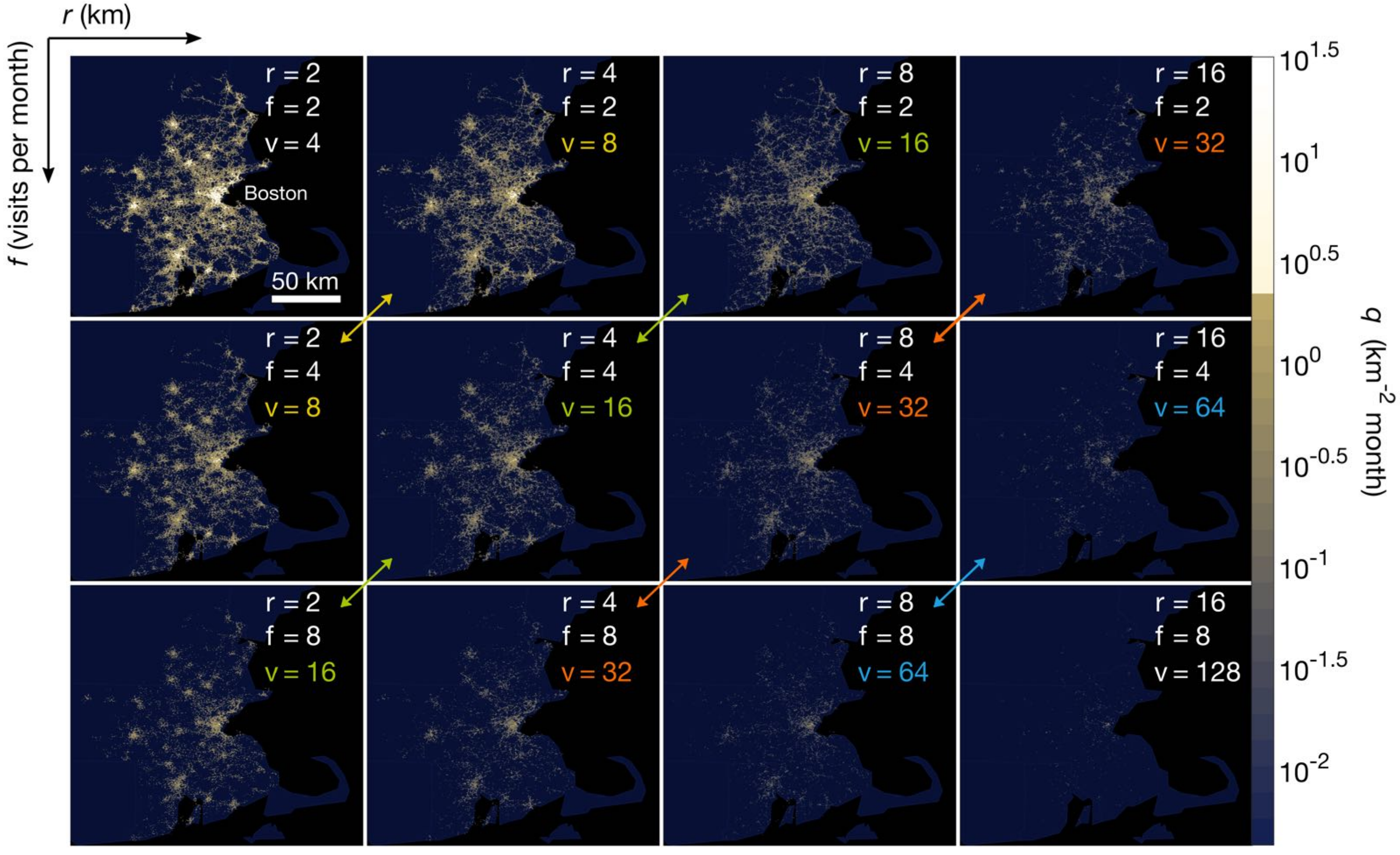}
\caption{\label{fig:satellite} {\footnotesize{\bf The spatio-temporal structure of movement in cities.} Panels show visitor influx maps for Greater Boston for different parameters $(r,f)$, as derived from geo-located mobile phone records (Fig.~\ref{fig:measurement} and Supplementary Information \mbox{section III}). The brightness of each grid cell ($500\,$m $\times$ $500\,$m) indicates the value of the visitor influx, $q$. Remarkably, visitor influx maps for the same visiting velocity $v=r\!f$ are nearly identical, as is clear from viewing along the diagonals indicated by the coloured arrows in the figure. Hence, doubling the visitation frequency $f$ (from top row to bottom row) results in the same quantitative decrease of the influx as doubling the travel distance $r$ (from left column to right column).}}
\end{figure}

To quantify both the spatial and temporal aspects of population flows, we define the flow $q$ to a location as the number of individuals travelling from a distance $r$ away and visiting with frequency $f$, per unit area of the origin and per unit frequency. For a fixed $f$, we observe that the flow decreases with distance $r$ (Fig.~\ref{fig:satellite}, from left to right). This generalises the well-known distance decay of social interactions -- for which however no general expression has been derived so far and which is typically reported without temporal considerations\cite{barbosa2018hm} -- to include distinct travel frequencies. Similarly, keeping $r$ fixed, $q$ decreases with increasing visitation frequency $f$ (Fig.~\ref{fig:satellite}, from top to bottom). 

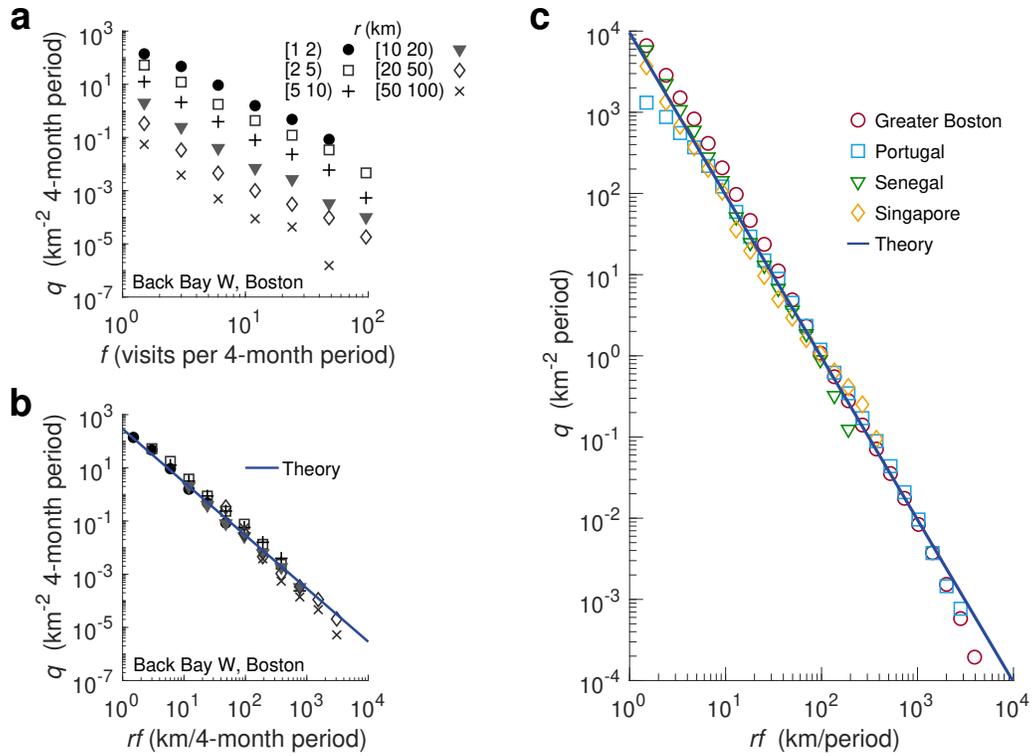
\begin{figure}
	\centering
	\begin{tikzpicture}
	\draw (0, 0) node[inner sep=0] {\includegraphics[width=0.8\textwidth]{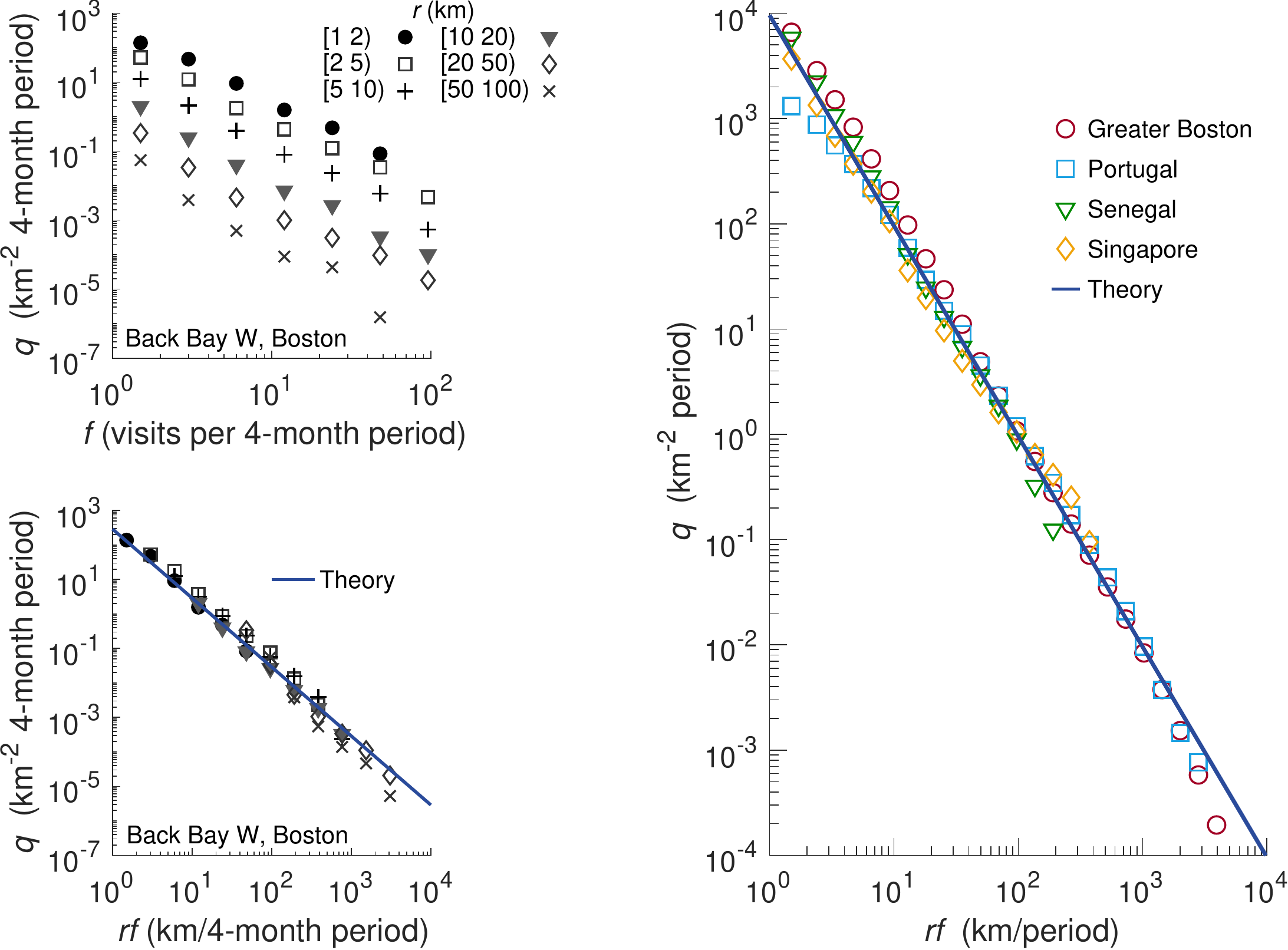}};
	\draw (-6.8, 4.9) node {\fontfamily{phv}\selectfont \textbf{\large a}};
	\draw (-6.8, -0.2) node {\fontfamily{phv}\selectfont \textbf{\large b}};
	\draw (0.1, 4.9) node {\fontfamily{phv}\selectfont \textbf{\large c}};
	\end{tikzpicture}
	\caption{\label{fig:collapse} {\footnotesize{\bf Universal scaling of population flows.} {\bf a}, Influx $q$ decreases with increasing frequency of visitation $f$ and travel distance $r$, as shown for the example of Back Bay West in Boston. {\bf b}, Rescaled values collapse onto a single curve, making the influx dependent only on the single variable $v=r\!f$. The entire frequency-distance spectrum is in striking agreement with the theoretically predicted scaling relation, $q \propto (rf)^{-2}$, shown as a straight line. The same pattern is generally obtained regardless of the specific location selected, as is clear from Figs.~6-12 and Supplementary Figs.~S6-S13. {\bf c}, Rescaled influx across Greater Boston, Portugal, Senegal and Singapore, demonstrating that the universal scaling holds for radically different urban regions worldwide. The underlying data are the empirical flows to a total of 20,672 locations (Supplementary Information section~IIIE). Symbols show the average values for each bin. Error bars (s.e.m.) are below symbol size. To visually compare the different world regions, we superimposed the shown curves by normalising the frequency-distance spectrum of each individual location with the average value  of the corresponding region \mbox{$\langle q \rangle = \langle F(rf) \rangle$} at \mbox{$ rf = 100 \, \text{km/period}$}. The inverse square of visiting velocity (straight line) describes the data very well.}}
\end{figure}

As a first step in developing our model for understanding this behaviour {\it quantitatively}, we use a dimensional argument to reduce the $(r,f)$ phase space of spatio-temporal flows to a single variable. In addition to $r$ and $f$, the flows $q$ to different locations typically depend on many diverse factors including the average travel speed in the city $u$, the characteristic length of the city $l$ (e.g., length of the roads) and a series of location-specific socio-economic factors $\{y_1,y_2,\ldots,y_n\}$ (e.g., population density, number of amenities)\cite{batty2013nsc}. Dimensional analysis, however, suggests that this dependence can be expressed mathematically in a dimensionless and simplified form (see derivation in the Supplementary Information section II) as $\tilde{q} = \tilde{F}\left(\frac{rf}{u},\left\{\frac{y_i}{y_n}\right\}_{1\leq i< n}\right)$, where $\tilde{q} \equiv ulq$ and $\tilde{F}$ is some unknown function. We therefore expect that for any fixed location $\mathbf{x}$ the dominant variation of $\tilde{q}_\mathbf{x}$ and therefore of $q_\mathbf{x}$ is determined by $rf$,
\begin{align}
q_\mathbf{x} = F_\mathbf{x}(rf) \label{eq:dimAnalysis}
\end{align}
where the function $F_\mathbf{x}$ remains to be determined. Indeed, any visitor whose home is at a distance $r$ covers, on average, a distance $rf$ per unit of time on the way towards the location $\mathbf{x}$. We therefore refer to the radial speed $v \equiv rf$ as the visiting velocity. It can be thought of as analogous to the bulk drift velocity of a gas driven by a pressure gradient and is to be distinguished from $u$, the average travel speed of an individual in the city. Our theoretical expectation is consistent with the data: the histogram of $q_\mathbf{x}$ in terms of the rescaled variable $rf$ shows a striking collapse of our measurements onto a single curve as illustrated in Fig.~\ref{fig:satellite} (compare the panels along the diagonals for $v=8$, $v=16$, $v=32$, and $v=64\,$km/month) and shown in more detail for the example of Back Bay West in Boston  (Fig.~\ref{fig:collapse}a,b). This same empirical validation is confirmed through the analysis of the flows to more than 20,000 locations in cities worldwide (Fig.~\ref{fig:collapse}c, Figs.~\ref{fig:indScalingGBI}-\ref{fig:indScalingSingI} and Supplementary Figs.~S6-S13), which reveals a hitherto neglected yet remarkable symmetry of human mobility: doubling the visitation frequency has exactly the same effect on the resulting flows as doubling the distance travelled. For instance, the number of individuals visiting once a month from \mbox{20$\,$km} away \mbox{($v=20\,$km/month)} equals the number of individuals visiting twice a month from \mbox{10$\,$km} away, or visiting \mbox{4 times} a month from 5$\,$km away, and so on -- in a similar way as a hypothesised collective `travel budget'\cite{lill1891rsa}. \textbf{This is our first main result: the flow to any location depends on the single variable $\bm{v=r\!f}$ and not on the variables $r$ and $f$ separately.}

Can we predict the mathematical form of $F_\mathbf{x}$? To answer this question, we start from the general observation that travellers typically seek the fastest and shortest route, one that takes the least time and traverses the shortest distance\cite{batty2013nsc}. Ideally, this means travelling along straight lines, which in a city is usually impossible. There is no choice but to follow the meandering roads and rail lines, so any specific journey involves a zigzagging route. However, when viewed at a larger scale with a coarse-grained spatio-temporal resolution, aggregated and averaged over all journeys and over long periods of time, we can assume a steady-state configuration in which a constant number of people are travelling approximately radially along spokes of circles at constant speed $v=rf$. The centre of each circle is their specific destination which acts as a hub.

To translate this behaviour into a mathematical model, let us define $Q_f(R)$ as the number of unique individuals who are visiting location $\mathbf{x}$ with frequency $f$ and whose home (considered permanent) is located within the disk $B(\mathbf{x},R)$ of radius $R$ around $\mathbf{x}$. We can then conceptualise the attracting destination as a `well' (or `magnet') that generates a `potential field' triggering the inward radial movement of visitors at speed $v=rf$. Since all trips are assumed to start from home (which is not restricted to commuting\cite{schneider2013udh}), the visitors can be modelled as individuals that remain located at home, but carry an attraction potential that is equivalent to the local flux $\bm{j}_f(r) := \rho_f(r)\bm{v}_f(r)$, where $\rho_f(r)$ is the local density of visitors and $\bm{v}_f(r)$ is the radial visiting velocity with $\|\bm{v}_f(r)\| = rf$ (Fig.~3a). Our main working hypothesis, again motivated by the hydrodynamic analogy of a well, is that for a fixed $\mathbf{x}$ the total attraction potential carried by a contour line (i.e. a circle around $\mathbf{x}$) is constant. More explicitly, we assume that for any $ r_1,r_2 \geq r_\varepsilon$, where $r_\varepsilon$ is the cut-off size of the location,
\begin{align}
\oint_{\dot{B}(\mathbf{x},r_1)}\bm{j}_f\cdot d\bm{\dot{B}} \approx \oint_{\dot{B}(\mathbf{x},r_2)}\bm{j}_f\cdot d\bm{\dot{B}}
\end{align}
where $\dot{B}(\mathbf{x},r)$ denotes the circle of centre $\mathbf{x}$ and radius $r$. This hypothesis is well supported by the data (Supplementary Information section~IVA). If we further assume that the flows are approximately radially symmetric (see Supplementary Information section~IVB for a detailed discussion), we get $2\pi r_1\rho_f(r_1)v_f(r_1) \approx 2\pi r_2\rho_f(r_1)v_f(r_2)$. In particular, for $r_1 = r_\varepsilon$ and any $r > r_\varepsilon$
, $\rho_f(r) \approx\rho_{f,\varepsilon}{r_\varepsilon^2}/{r^2} \propto {1}/{r^2}$.

We can now deduce that the number of individuals visiting with frequency $f$ from $B(\mathbf{x},R)$ verifies
\begin{align}
\begin{aligned}
Q_f(R) & := \iint_{B(\mathbf{x},R)} \rho_f(r) \ dB\\
       & \propto \ln\left(\frac{R}{r_\epsilon}\right)
\end{aligned}
\end{align}
Data from across the globe strongly support the resulting prediction that the total number of visitors across all frequencies increases logarithmically with distance $R$, $Q^\text{tot}(R) := \sum_f Q_f(R) \propto\ln(R/r_\varepsilon)$ (Fig. 3b,c). Finally, we can also deduce that the number of visitors per frequency and per area to location $\mathbf{x}$, $q_\mathbf{x} \coloneq \frac{1}{2 \pi r} \frac{\partial^2 Q_\mathbf{x}^{\text{tot}}}{\partial r \partial f}$ must solve \mbox{$\int_{f_\text{min}}^{f_\text{max}} q_\mathbf{x}(rf)df \propto 1/r^2$}. Our theory therefore predicts
\begin{equation}
q_\mathbf{x} \propto (rf)^{-2}
\end{equation}

Empirical population flows in different urban regions across the world are in excellent agreement with the theoretically predicted inverse square law (Fig.~\ref{fig:collapse}b,c, Figs.~6-12 and Supplementary Information sections~IVC-D). Importantly, the discovered regularity is largely unaffected by location-specific conditions, including strong variations in surrounding population densities or in the level of economic or infrastructural development, and thus signals a universal principle behind the organisation of radically different urban systems. \textbf{This is our second main result: the flow of individuals to any given location scales as the inverse square of visiting velocity.}
 
The discovered scaling relation implies that for any location the magnitude of the entire frequency-distance spectrum of flows can be obtained by just knowing one single point on the scaling curve. This simplification opens up an unprecedented range of possibilities for the prediction of various flow quantities between each pair of locations. To demonstrate this, we present a simple method that uses population density $\rho$ as the sole input. The location-specific proportionality constant $k_\mathbf{x}$ such that $q_\mathbf{x}(rf)=k_\mathbf{x}/(rf)^2$ can be estimated by
assuming that individuals return back home on a daily basis\cite{barbosa2018hm}, which gives rise to a local flow with minimum frequency $f_{\text{home}} \approx 1 \, \text{day}^{-1}$. Using our continuous hydrodynamic analogy, this can be translated into a  boundary condition at $r=r_\varepsilon$, dictating that the minimum visiting frequency of all individuals living directly on the boundary of an attracting location corresponds to the minimum frequency of all individuals living just inside the boundary (Fig.~\ref{fig:model_boundary_condition}). This can be written as $\rho(r_\varepsilon)=\frac{\partial Q_\mathbf{x}^{\text{tot}}}{2\pi r \partial r }\Bigr|_{\substack{r=r_\varepsilon}} = \int_{f_{\text{home}}}^{\infty} q_\mathbf{x}(r_\varepsilon f) \mathrm{d}f$, so that \mbox{$k_\mathbf{x} = \rho(r_\varepsilon) r_\varepsilon^2 f_{\text{home}}$}. Once the value of $k_\mathbf{x}$ has been evaluated, the average daily number of trips from location $\mathbf{x'}$ to location $\mathbf{x}$ is simply given by $V_{\mathbf{x'},\mathbf{x}} \approx \mathcal{A}_\mathbf{x'} \! \int_{f_\text{min}}^{f_\text{max}} f q_\mathbf{x} \ud f  = k_\mathbf{x} \mathcal{A}_\mathbf{x'} /r^2 \ln \left(  f_{\text{max}} / f_\text{min} \right)$ with $\mathcal{A}_\mathbf{x'}$ being the area of the origin location and $r = \| \mathbf{x} - \mathbf{x'} \|$. The frequency limits can either be set according to the objective of the study (e.g., if the focus is on high-frequency visitors) or, if all types of trips are to be included, they can be set as $f_\text{min} = 1 / T$, where $T \gg 1 \,  \text{day}$ is the observation period, and $f_\text{max} = 1 \, \text{day}^{-1}$ (for $r \gg  r_\varepsilon$). Because of the logarithmic form, even a large error in the estimation of the frequency limits does not appreciably affect $V_{\mathbf{x'},\mathbf{x}}$. The number of visiting individuals, $Q_{\mathbf{x'},\mathbf{x}} \approx \mathcal{A}_{\mathbf{x'}}  \int \! q_\mathbf{x} \ud f $, is obtained in the same manner. All of these predictions are in remarkable agreement with the data (Fig.~\ref{fig:prediction}). Besides population density, our framework allows for the use of a broad range of alternative input quantities (Supplementary Information section~IVE). For instance, if the total rate of trips arriving at a given location is known (e.g., from simple traffic counts), the flows between individual locations can be calculated for different times of the day, without the need to specify the frequency limits (Supplementary Information section~IVF).

We further compare these predictions to those of the gravity model\cite{zipf1946phi} and the radiation model\cite{simini2012umm}, which are currently the most widely used mobility models for the flow of individuals\cite{barbosa2018hm} (Supplementary Information section VIA). These existing approaches have been criticised for their lack of predictive power at fine-grained (intra-urban) spatial scales, which shows that they miss some of the essential mechanisms behind the population flows inside cities\cite{masucci2013gvr,yang2014lpc}. This is indeed confirmed by our analysis (Fig.~\ref{fig:prediction}a). Another fundamental limitation is that they do not take into account the frequency of flows. As a consequence, the gravity model requires separate parameter calibrations for each specific flow quantity so that knowing the number of {\it trips} does not allow to infer how many {\it unique individuals} are actually visiting a location over time and vice-versa (Fig.~\ref{fig:prediction}b,c and Supplementary Information section VIA). Similarly, since each individual may give rise to several trips to multiple locations, the radiation model can predict the number of {trips} (Fig.~\ref{fig:prediction}a) but not the number of {unique individuals} that are visiting a location over time for any purpose (Fig.~\ref{fig:prediction}b,c and Supplementary Information section VIA). This oversimplification of the heterogeneity of travel frequencies through indistinct aggregation in such established models may lead to biased conclusions regarding the spatial mixing of people with potentially far-reaching consequences \mbox{(e.g., for} the modelling of infectious diseases). Our theory resolves both of these fundamental limitations. It is based on general principles that are applicable in particular to fine-grained spatial scales (Fig.~\ref{fig:prediction}a) and it allows for the simultaneous prediction of both the number of trips {\it and} the number of individuals across the entire frequency spectrum without the need for model calibrations (Fig.~\ref{fig:prediction}b,c). This superior performance of our framework is empirically confirmed in all the world regions considered \mbox{here (Fig.~\ref{fig:prediction}d)}.


Several additional consequences follow from our theory. First, it makes it possible to accurately predict the total number of miles travelled that a particular location induces over time (e.g., for estimating traffic-related CO$_{\text 2}$ emissions), to determine a location's range of influence (e.g., for the planning of infrastructures or business establishments) or to explore social mixing patterns. Second, and perhaps surprisingly, the derived scaling relation also implies that, on average, the distance travelled {\it per visit} does not depend on the overall attractiveness of a location (in terms of the total number of visits it receives, Supplementary Information section~IVG). Deviations from this baseline allow for the identification of unusual hotspots of activity (Supplementary Information section~IVH) or locations in need of stimulation. {Third, as is shown in the follow-up work of Dong et al.\cite{dong2020spectral}, the discovered scaling law is consistent with a fine-grained mechanistic model that explicitly considers the choices of individuals. More precisely, this model extends the well known `exploration and preferential return model' of Song et al.\cite{song2010msp} by modulating the probability of an individual visiting a given location based on its popularity. It is shown that this individual-based model is indeed an explicit realization of the universal scaling law derived and discovered in the current paper.}	
	
Given the extensive literature on and detailed analyses of movement and transport in cities, it is surprising that the simple but powerful result that we have derived here had not yet been discovered. It constitutes a general quantitative constraint on one of the most fundamental aspects of socio-economic life that is independent of geography, culture and degree of development. Furthermore, it is well supported by an unprecedented data set that is representative of urban systems worldwide. Clearly, further investigation needs to be forthcoming to test the boundaries of these predictions and to develop the theory in terms of its relationship to the underlying infrastructural and social network dynamics that constitute urban life\cite{jacobs1961ld}. At the same time, the success of our parsimonious model in reproducing population flows has potentially important implications for practitioners dealing with the challenging problem of human mobility in our rapidly growing cities. It also makes our theory a suitable starting point for the study of collective movement patterns in domains beyond the ones considered here. 

\nolinenumbers

\normalsize

\bibliographystyle{naturemag}
\bibliography{references}

\begin{addendum}

\item[Acknowledgements\hspace{-0.25cm}]We thank Lu\'is M.A. Bettencourt, Roberta Sinatra and Pieter Herthogs for helpful discussions, Sebastian Grauwin for providing the Matlab code for the radiation model and Sarah Wicki for proofreading the manuscript. We acknowledge Airsage, ORANGE / SONATEL and Singtel for providing the data. This work was supported by the National Science Foundation, the AT\&T Foundation, the MIT SMART program, the MIT CCES program, Audi Volkswagen, BBVA, Ericsson, Ferrovial, GE, the MIT Senseable City Lab Consortium, the John Templeton Foundation (grant no. 15705), the Eugene and Clare Thaw Charitable Trust, the US Army Research Office Minerva Programme (grant no. W911NF-12-1-0097) and the Singapore National Research Foundation (FI 370074016).

\item[Author Contributions\hspace{-0.22cm}] M.Sc., M.S., C.R. and G.B.W. designed the study. M.Sc. processed and analysed the data. G.B.W. developed the theory. M.Sc., H.S. and G.B.W.  derived the model. All authors  discussed the results and wrote the manuscript.

\end{addendum}

\newpage
\clearpage
\thispagestyle{empty}

\clearpage
\thispagestyle{empty}
\clearpage
\thispagestyle{empty}
\begin{figure}
\centering
\begin{tikzpicture}
    \draw (0, 0) node[inner sep=0] {\includegraphics[width=0.75\textwidth]{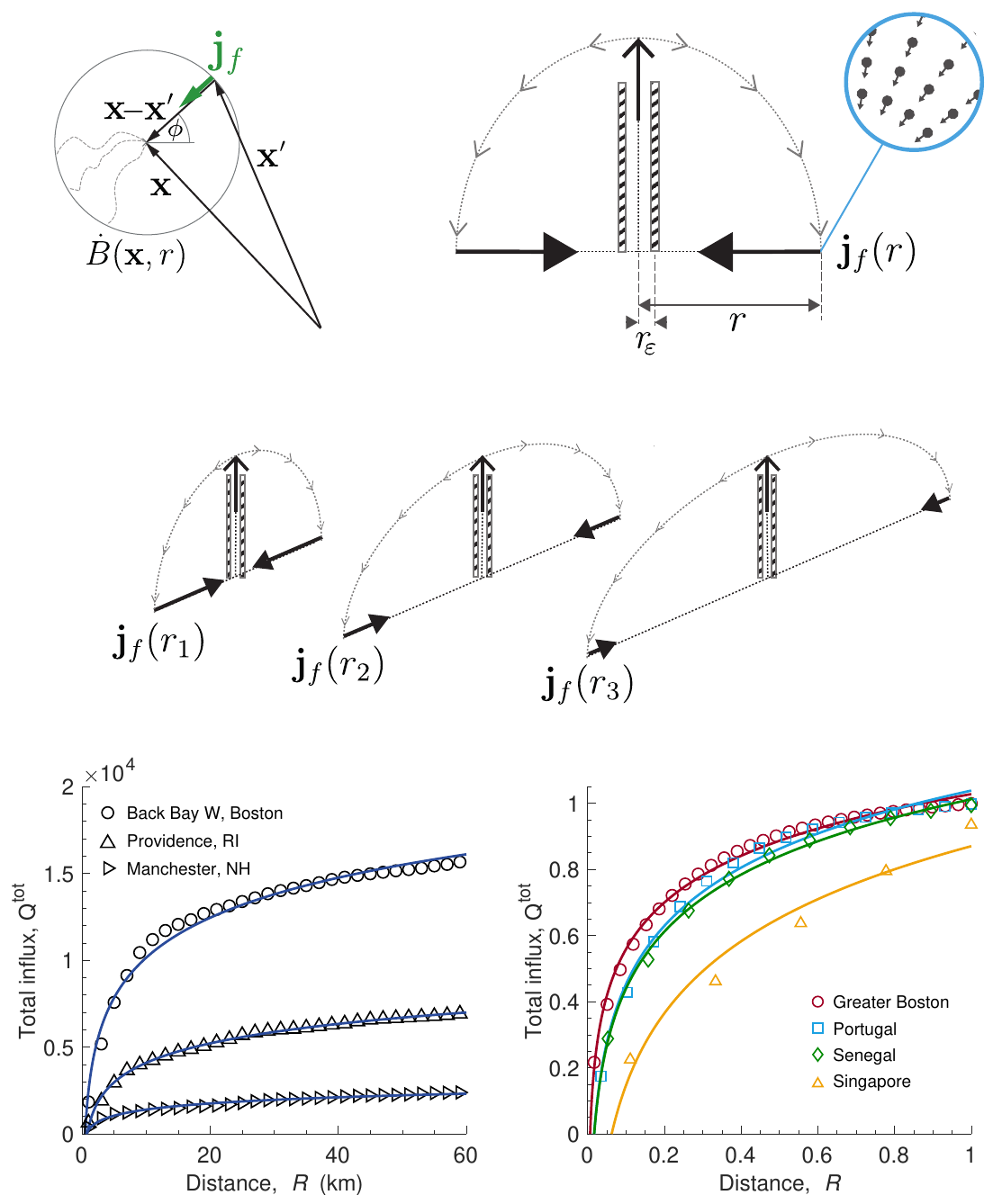}};
    \draw (-6.2, 7.2) node {\fontfamily{phv}\selectfont \textbf{\large a}};
    \draw (-6.2, -2.2) node {\fontfamily{phv}\selectfont \textbf{\large b}};
    \draw (0.2, -2.2) node {\fontfamily{phv}\selectfont \textbf{\large c}};
\end{tikzpicture}
\caption{{\footnotesize{\bf A simple model for scaling of visitor influx.} {\bf a}, Definition of the basic quantities of our flow model that is inspired by the hydrodynamic analogy of a well. Averaging over all meandering paths (top left, dashed lines) is assumed to result in a straight line for the collective flow to the attracting location (the \enquote*{well}) at $\mathbf{x}$ (green arrow). The location-specific total attraction potential (top right, indicated by the vertical arrow) results in a circular and radially symmetric local low $\mathbf{j}_f$ converging to $\mathbf{x}$. The dotted lines with arrows symbolise the symmetric flux back to the home location at distance $r = \| \mathbf{x} - \mathbf{x'} \|$. In direct analogy to a well, the same total attraction potential applies to all circles $\dot{B}(\mathbf{x},r)$ of centre $\mathbf{x}$ and any radius $r$. This model predicts a decrease of $\| \mathbf{j}_f(r) \|$ with increasing value of $r$ (bottom, from left to right) such that the total number of visitors $Q^{\text{tot}}$ increases logarithmically with $R$. {\bf b},{\bf c}, Empirical movement patterns across the globe are in excellent agreement with this theoretical prediction. {\bf b}, Total number of visitors versus distance for exemplary locations in Greater Boston. Lines are logarithmic fits (least-squares regression) to the data. {\bf c}, Total number of visitors across all locations in Greater Boston, Portugal, Senegal and Singapore, showing that the logarithmic increase holds for all urban regions studied here. Symbols are averages over all individual locations in each region, whereas both $Q^{\text{tot}}$ and $R$ are normalised by their maximum values.}}\label{fig:model}
\end{figure}
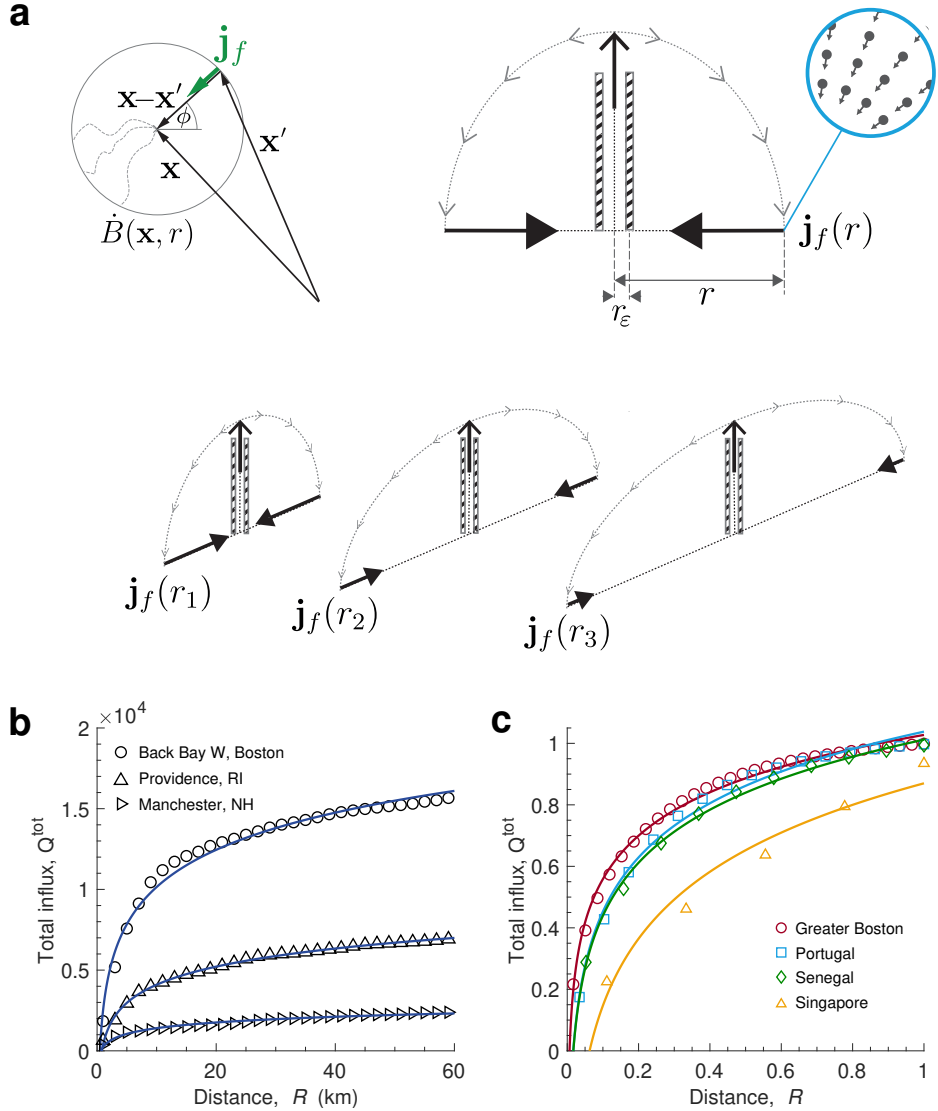

\clearpage
\thispagestyle{empty}
\begin{figure}
\centering
\begin{tikzpicture}
    \draw (0, 0) node[inner sep=0] {\includegraphics[width=0.85\textwidth]{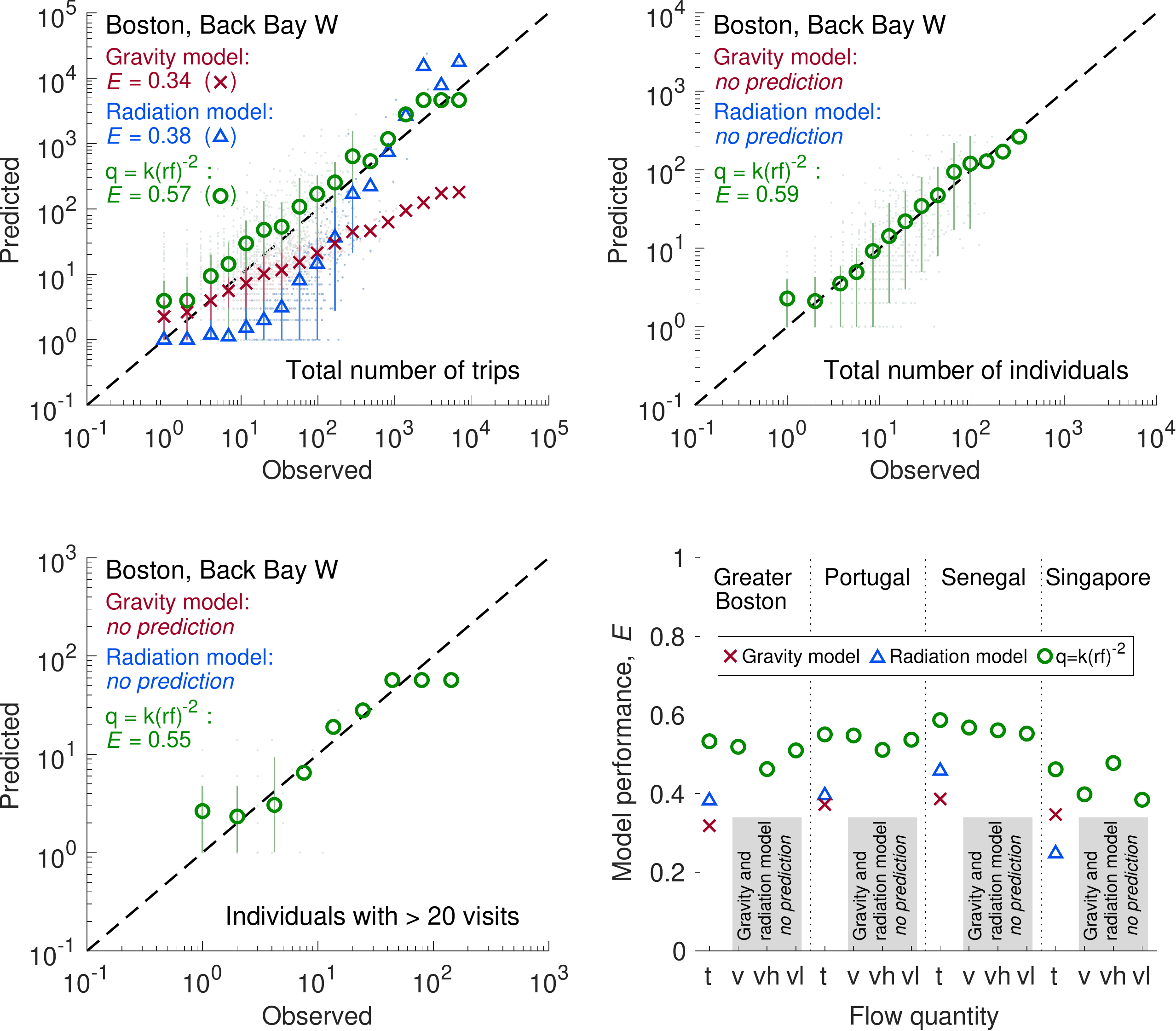}};
    \draw (-7.1, 6.2) node {\fontfamily{phv}\selectfont \textbf{\large a}};
    \draw (0.2, 6.2) node {\fontfamily{phv}\selectfont \textbf{\large b}};
    \draw (-7.1, -0.3) node {\fontfamily{phv}\selectfont \textbf{\large c}};
    \draw (0.2, -0.3) node {\fontfamily{phv}\selectfont \textbf{\large d}};
\end{tikzpicture}
\caption{\footnotesize{\bf Predicting the flows between individual locations.} {\bf a}, Predictions for the observed trips to Back Bay West, Boston, derived from state-of-the-art models (the gravity law and the radiation model, described in Supplementary Information section~VIB) compared to predictions from our $rf$ scaling theory. Symbols are mean values for each bin and lines are the 0.1-0.9 quantiles. The dashed line corresponds to a perfect agreement between the observed values and the predictions, clearly showing that our theory systematically outperforms the existing models. The performance of each model is further quantified based on the S\o rensen-Dice similarity index ($E$), with $E = 1$ if there is a perfect match and $E=0$ if there is no match at all (Supplementary Information section VIB). {\bf b,} Number of (individual) visitors. The fitting parameters of the gravity law from the number of trips (a) do not allow to predict the number of individuals. The radiation model does not provide a prediction of the number of visitors either, because it assigns only one destination location to each individual. It is therefore unable to explicitly consider the fact that an individual may visit several different locations. {\bf c,} Number of high-frequency visitors. {\bf d,} Systematic comparison over all considered locations in the studied world regions ($\approx 2.9 \cdot 10^6$ origin-destination pairs in total, Supplementary Information section~VIB) for number of trips (t), number of visitors (v), number of high-frequency visitors (vh) and number of low-frequency visitors (vl). The gravity model  (calibrated for t) and the radiation model are unable to predict v, vh, or vl. Our theory overcomes this limitation.}\label{fig:prediction}
\end{figure}
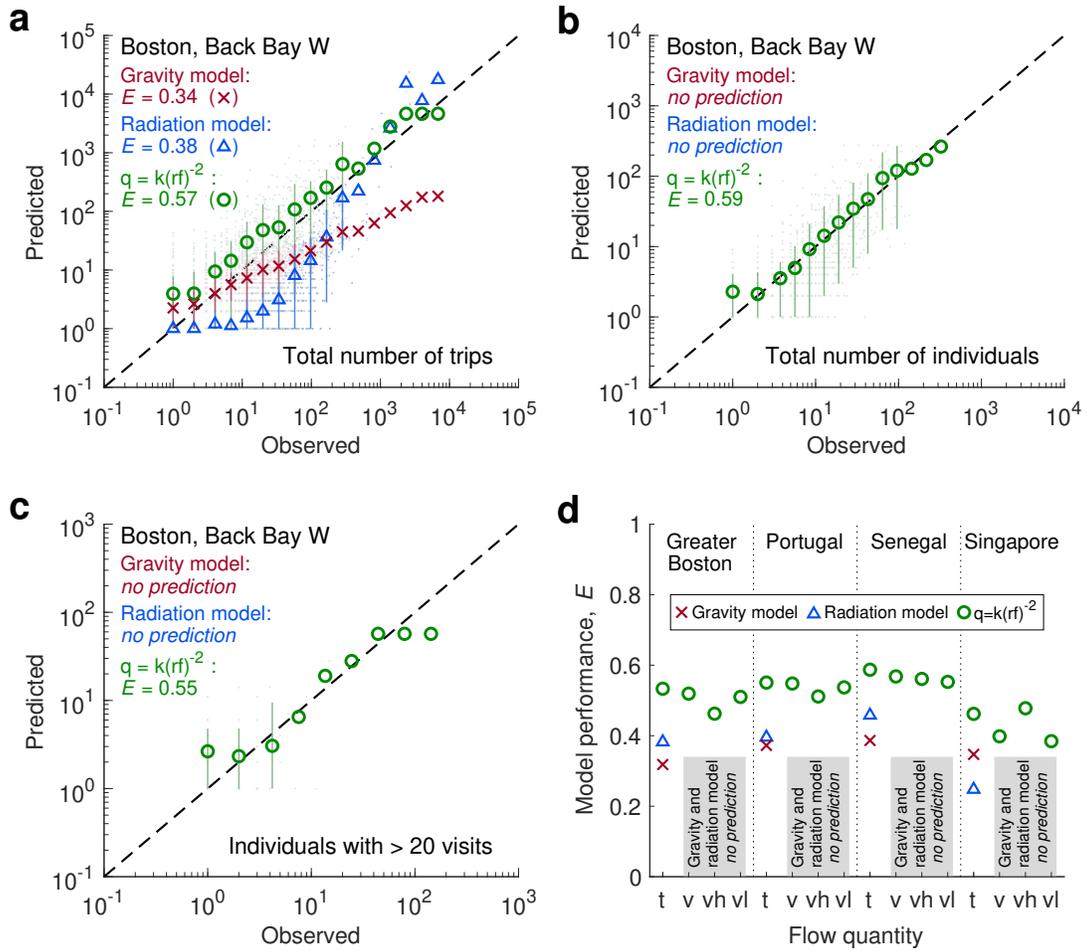


\clearpage
\thispagestyle{empty}
\begin{figure}
\centering\hspace{1cm}
\includegraphics[width=0.9\textwidth]{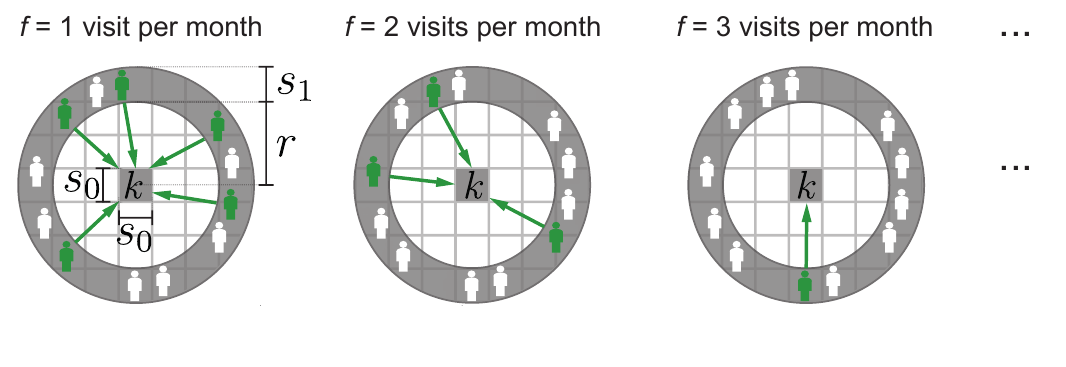}
\caption{\label{fig:measurement}{\footnotesize{\bf Measuring the empirical flow of individuals to locations}. To measure the flow $q$, we partition the regions into a regular grid with square cells of equal size $s_0 \times s_0$ For a grid cell $k$ we count the number of individuals who are living in a surrounding annulus with inner radius $r$ and width $s_1$, and who are visiting the cell with frequency $[f,f+\Delta f)$ (green coloured symbols, shown for exemplary values of $f$). The visitation frequency is defined here as the number of distinct days per month, during which a user is found in the grid cell (Supplementary Information section~IIIE). For the frequency bins we apply a fixed value of $\Delta f = 1\,$visit per day. In the main text, we use \mbox{$s_0 = 500\,$m} for Greater Boston and Singapore and $s_0 = 1\,$km for Portugal and Senegal (Supplementary Information section~IIIB). To demonstrate the robustness of our results, we additionally vary the cell size in the range $250\,{\rm m} \leq s_0 \leq 2\,{\rm km}$ (Supplementary Information section~V). We quantify the influx as $q_k \approx \Delta {Q}^{\rm tot}_{k} / (\Delta A \Delta f)$, where $\Delta {Q}^{\rm tot}_{k}$ is the count of users who live at distance $[r, r+s_1)$ and visit cell $k$ with frequency $[f,f+\Delta f)$, and $\Delta A$ is the area of the corresponding annulus with radius $r$ and width $s_1$. In the main text we use $s_1 = 1\,$km for all regions, while our results are robust against changes of this value (Supplementary Information section~V).}}
\end{figure}

\clearpage
\thispagestyle{empty}
\begin{figure}
\centering
\includegraphics[width=0.95\textwidth]{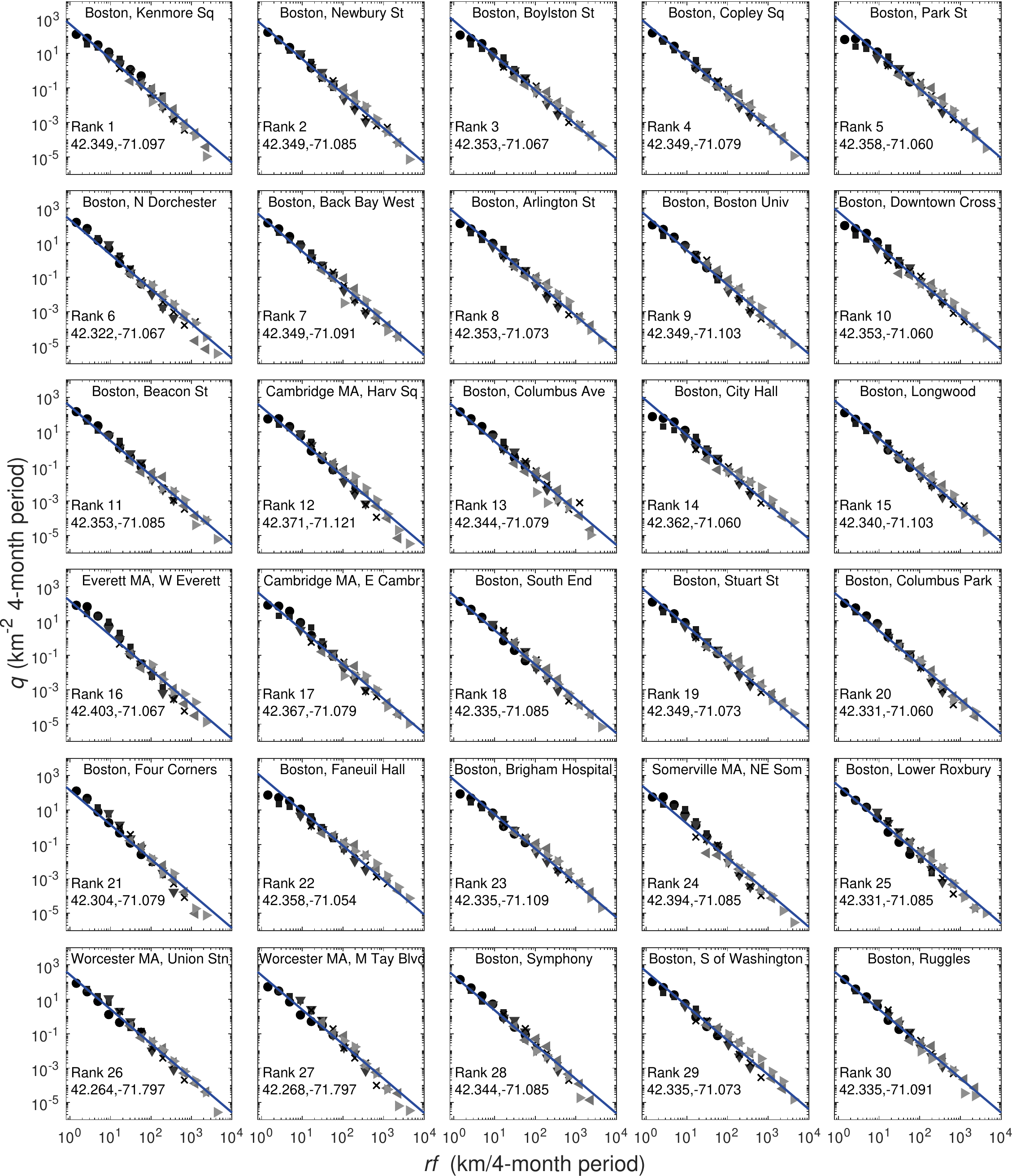}\vspace{-0.2cm}
\caption{\label{fig:indScalingGBI}{\footnotesize{\bf Universality of the scaling relation \boldmath${q \propto (rf)^{-2}}$ across Greater Boston.} The panels depict the data for individual locations (500$\,$m $\times$ 500$\,$m grid cells), ranked according to the total number of visitors from neighbouring cells. Shown are locations of rank 1-30 (from top left to bottom right). Symbols denote different frequency bins, with \protect\markerone~for $[1,2)$, \protect\markertwo~for $[2,5)$, \protect\markerthree~for $[5,10)$, $\bm{\times}$ for $[10,20)$, \protect\markerfive~for $[20,50)$ and \protect\markersix~for $[50,100)$ visits per 4-month period. The geographic coordinates of each location (latitude and longitude of the centre point of the grid cell) are indicated. The straight lines denote the theoretically predicted inverse square of $rf$.}}
\end{figure}

\clearpage
\thispagestyle{empty}
\begin{figure}
\centering
\includegraphics[width=0.95\textwidth]{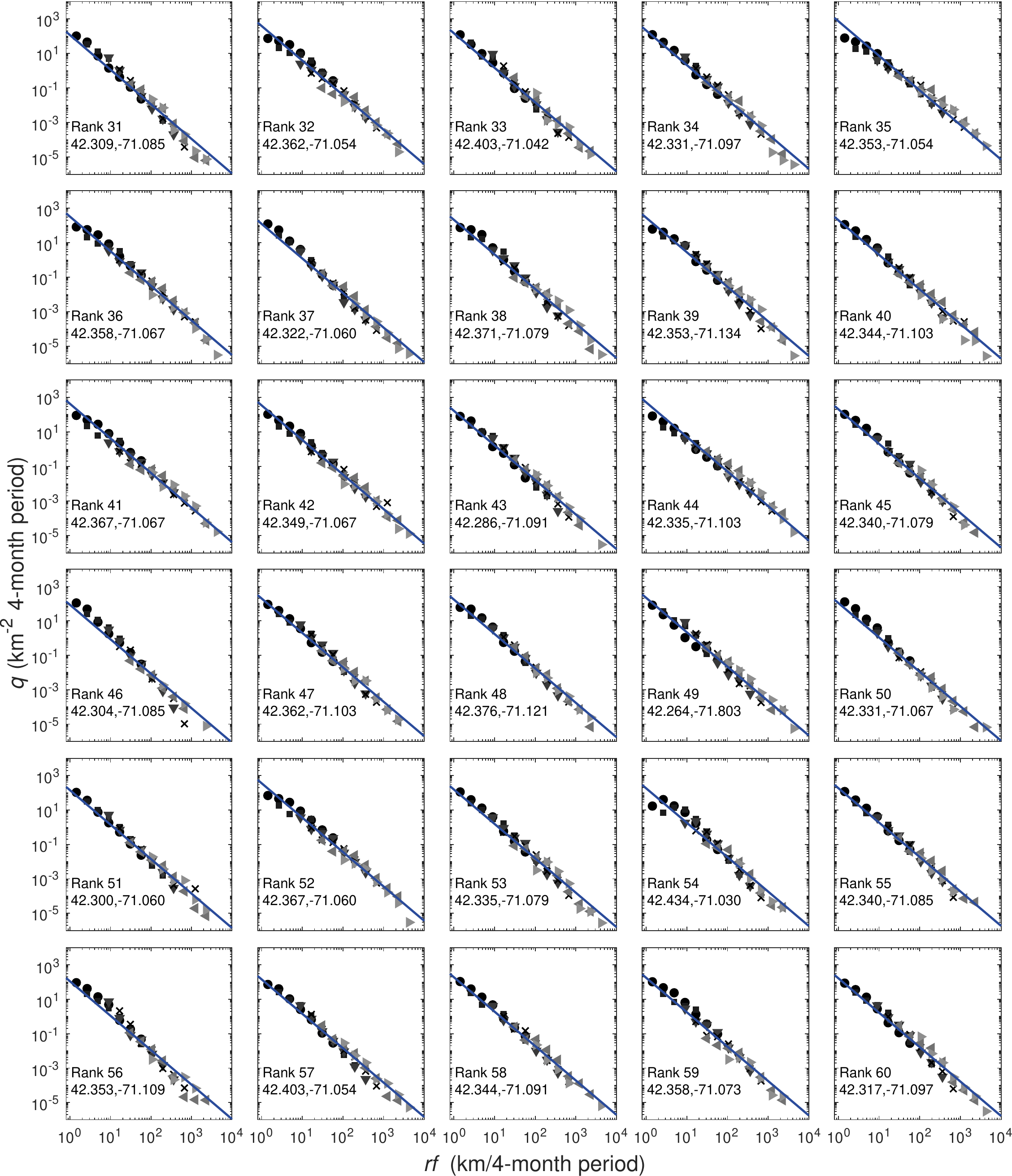}\vspace{-0.2cm}
\caption{\label{fig:indScalingGBII}{\footnotesize{\bf Universality of the scaling relation \boldmath${q \propto (rf)^{-2}}$ across Greater Boston.} The panels depict the data for individual locations (500$\,$m $\times$ 500$\,$m grid cells), ranked according to the total number of visitors from neighbouring cells. Shown are locations of rank 31-60 (from top left to bottom right). Symbols and notations are as in \mbox{Fig.~\ref{fig:indScalingGBI}}.}}
\end{figure}

\clearpage
\thispagestyle{empty}
\begin{figure}
\centering
\includegraphics[width=0.95\textwidth]{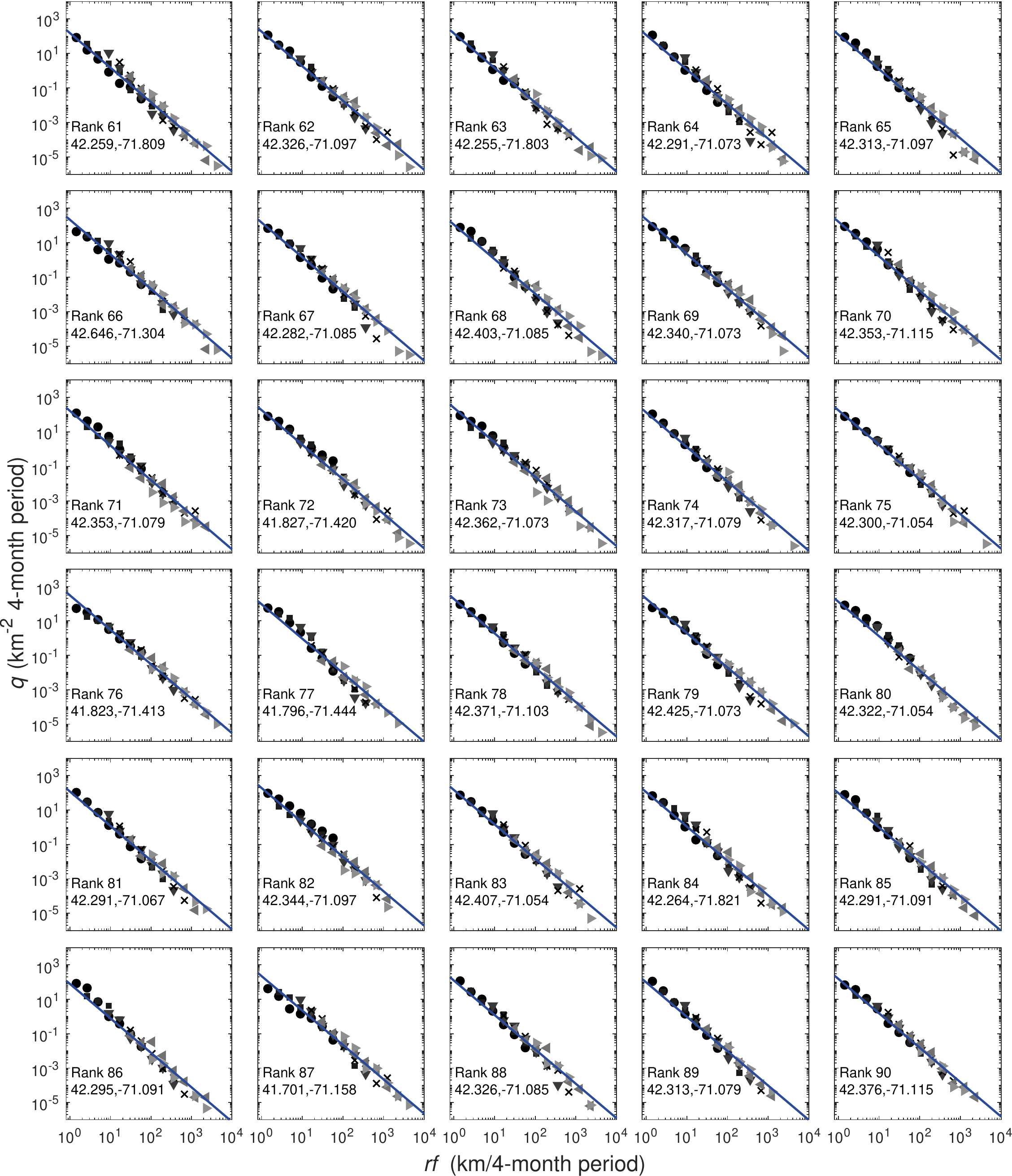}\vspace{-0.2cm}
\caption{\label{fig:indScalingGBIII}{\footnotesize{\bf Universality of the scaling relation \boldmath${q \propto (rf)^{-2}}$ across Greater Boston.} The panels depict the data for individual locations (500$\,$m $\times$ 500$\,$m grid cells), ranked according to the total number of visitors from neighbouring cells. Shown are locations of rank 61-90 (from top left to bottom right). Symbols and notations are as in \mbox{Fig.~\ref{fig:indScalingGBI}}.}}
\end{figure}

\clearpage
\thispagestyle{empty}
\begin{figure}
\centering
\includegraphics[width=0.95\textwidth]{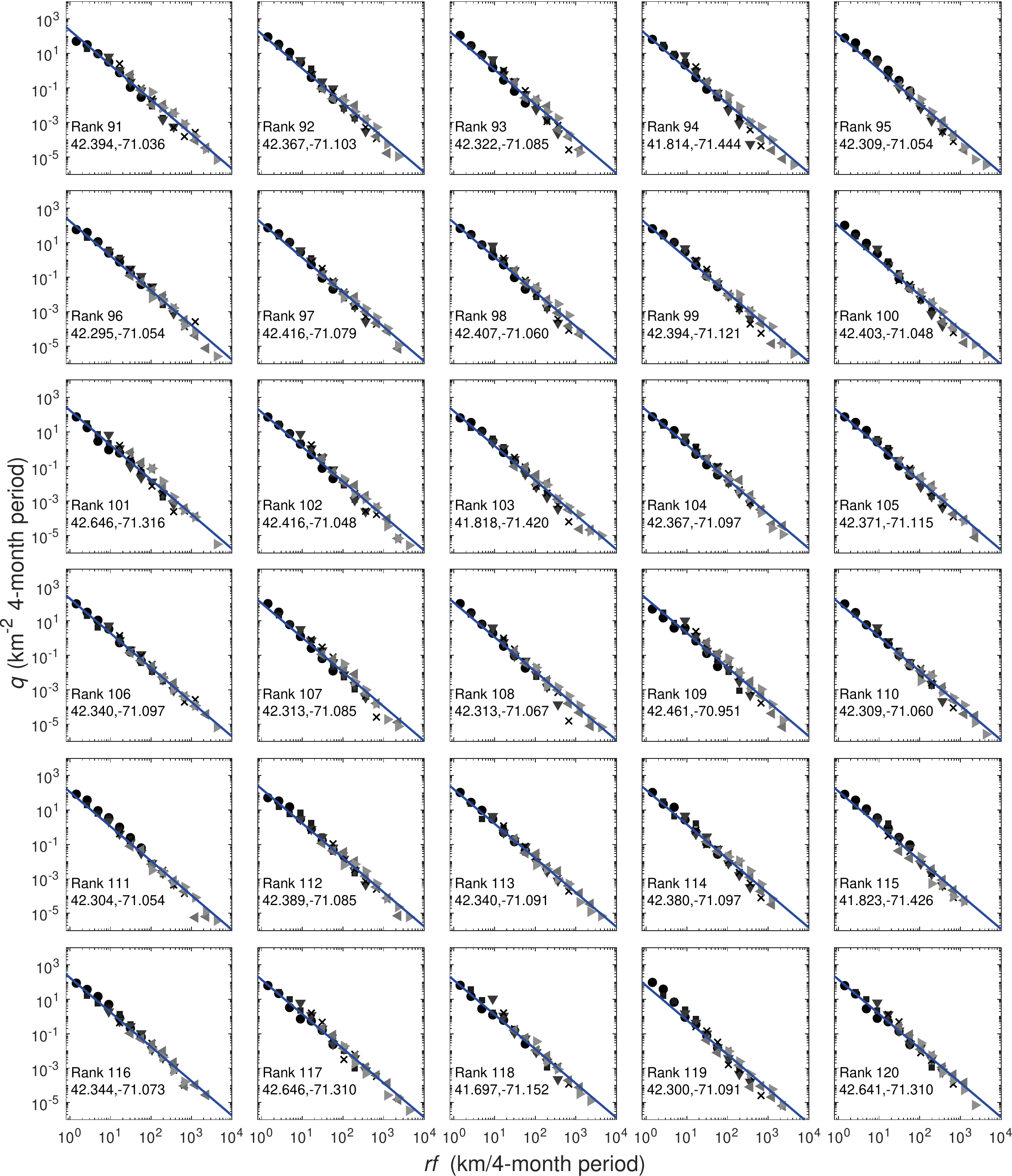}\vspace{-0.2cm}
\caption{\label{fig:indScalingGBIV}{\footnotesize{\bf Universality of the scaling relation \boldmath${q \propto (rf)^{-2}}$ across Greater Boston.} The panels depict the data for individual locations (500$\,$m $\times$ 500$\,$m grid cells), ranked according to the total number of visitors from neighbouring cells. Shown are locations of rank 91-120 (from top left to bottom right). Symbols and notations are as in \mbox{Fig.~\ref{fig:indScalingGBI}}.}}
\end{figure}

\clearpage
\thispagestyle{empty}
\begin{figure}
\centering
\includegraphics[width=0.95\textwidth]{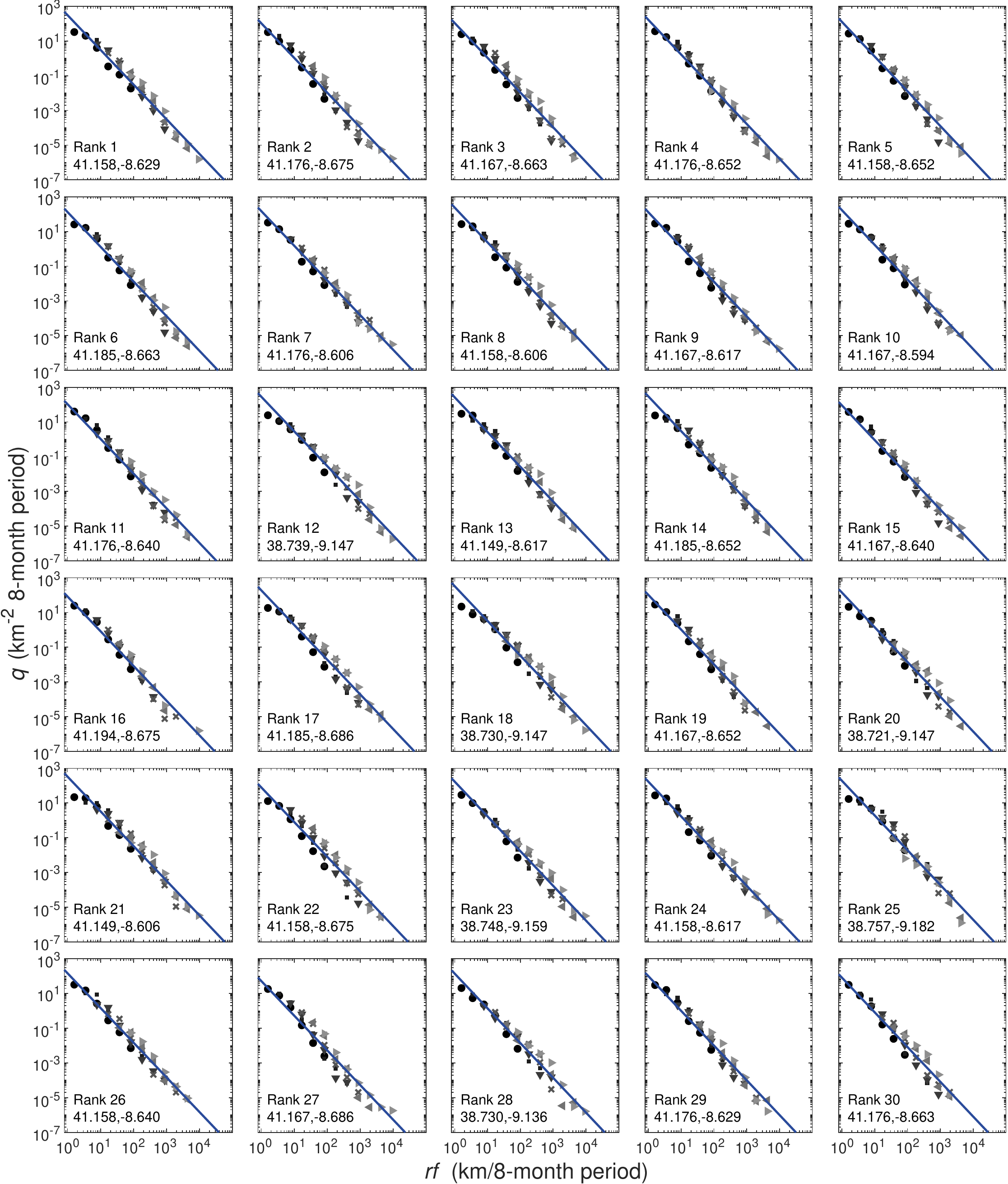}\vspace{-0.2cm}
\caption{\label{fig:indScalingPI}{\footnotesize{\bf Universality of the scaling relation \boldmath${q \propto (rf)^{-2}}$ across Portugal.} The panels depict the data for individual locations (1$\,$km $\times$ 1$\,$km grid cells), ranked according to the total number of visitors from neighbouring cells. Shown are locations of rank 1-30 (from top left to bottom right). Symbols denote different frequency bins, with \protect\markerone~for $[1,2)$, \protect\markertwo~for $[2,5)$, \protect\markerthree~for $[5,10)$, $\bm{\times}$ for $[10,20)$, \protect\markerfive~for $[20,50)$ and \protect\markersix~for $[50,100)$ visits per 8-month period. The geographic coordinates of each location (latitude and longitude of the centre point of the grid cell) are indicated. The straight lines denote the theoretically predicted inverse square of $rf$.}}
\end{figure}

\clearpage
\thispagestyle{empty}
\begin{figure}
\centering
\includegraphics[width=0.95\textwidth]{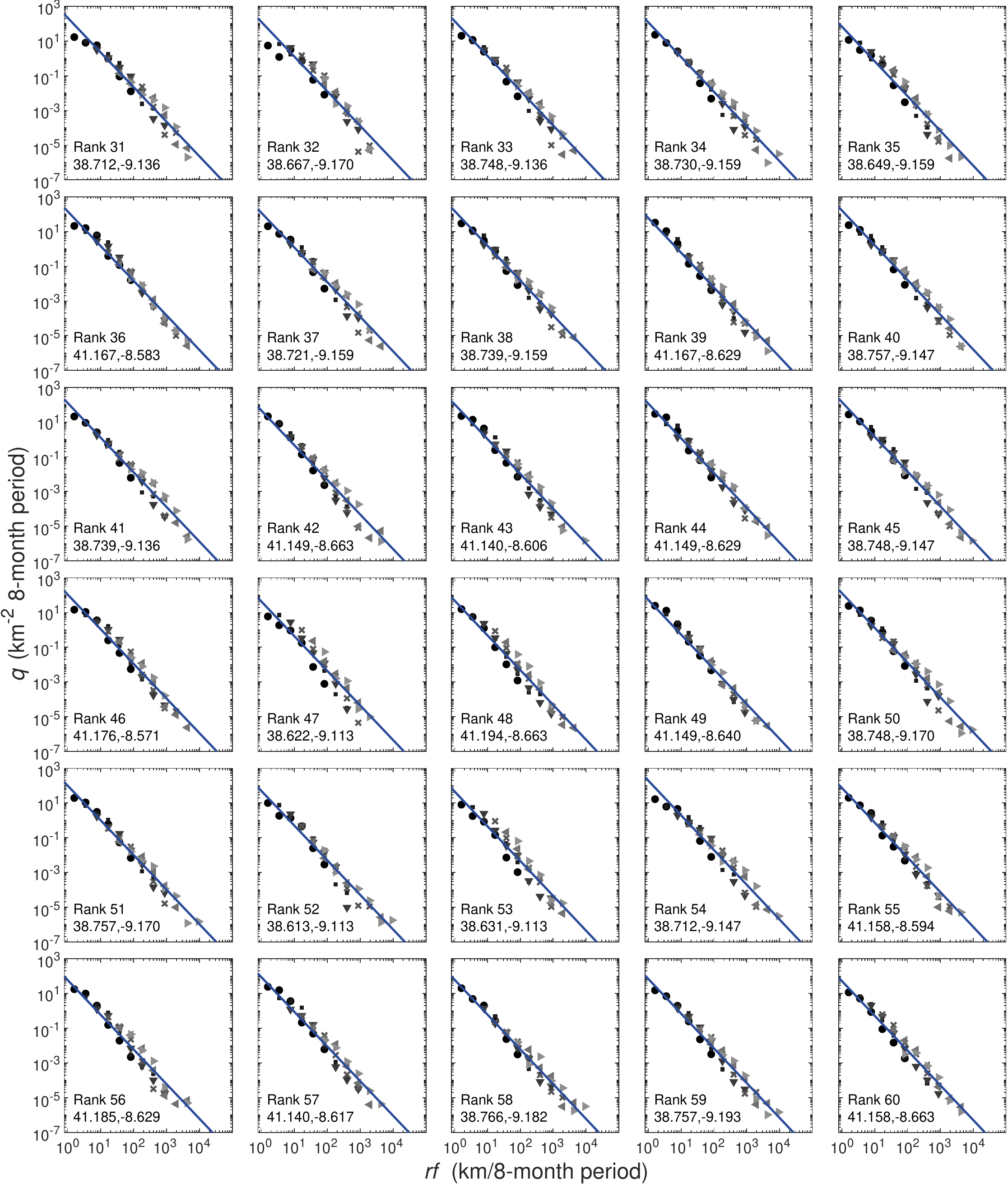}\vspace{-0.2cm}
\caption{\label{fig:indScalingPII}{\footnotesize{\bf Universality of the scaling relation \boldmath${q \propto (rf)^{-2}}$ across Portugal.} The panels depict the data for individual locations (1$\,$km $\times$ 1$\,$km grid cells), ranked according to the total number of visitors from neighbouring cells. Shown are locations of rank 31-60 (from top left to bottom right). Symbols and notations are as in \mbox{Fig.~\ref{fig:indScalingPI}}.}}
\end{figure}

\clearpage
\thispagestyle{empty}
\begin{figure}
\centering
\includegraphics[width=0.95\textwidth]{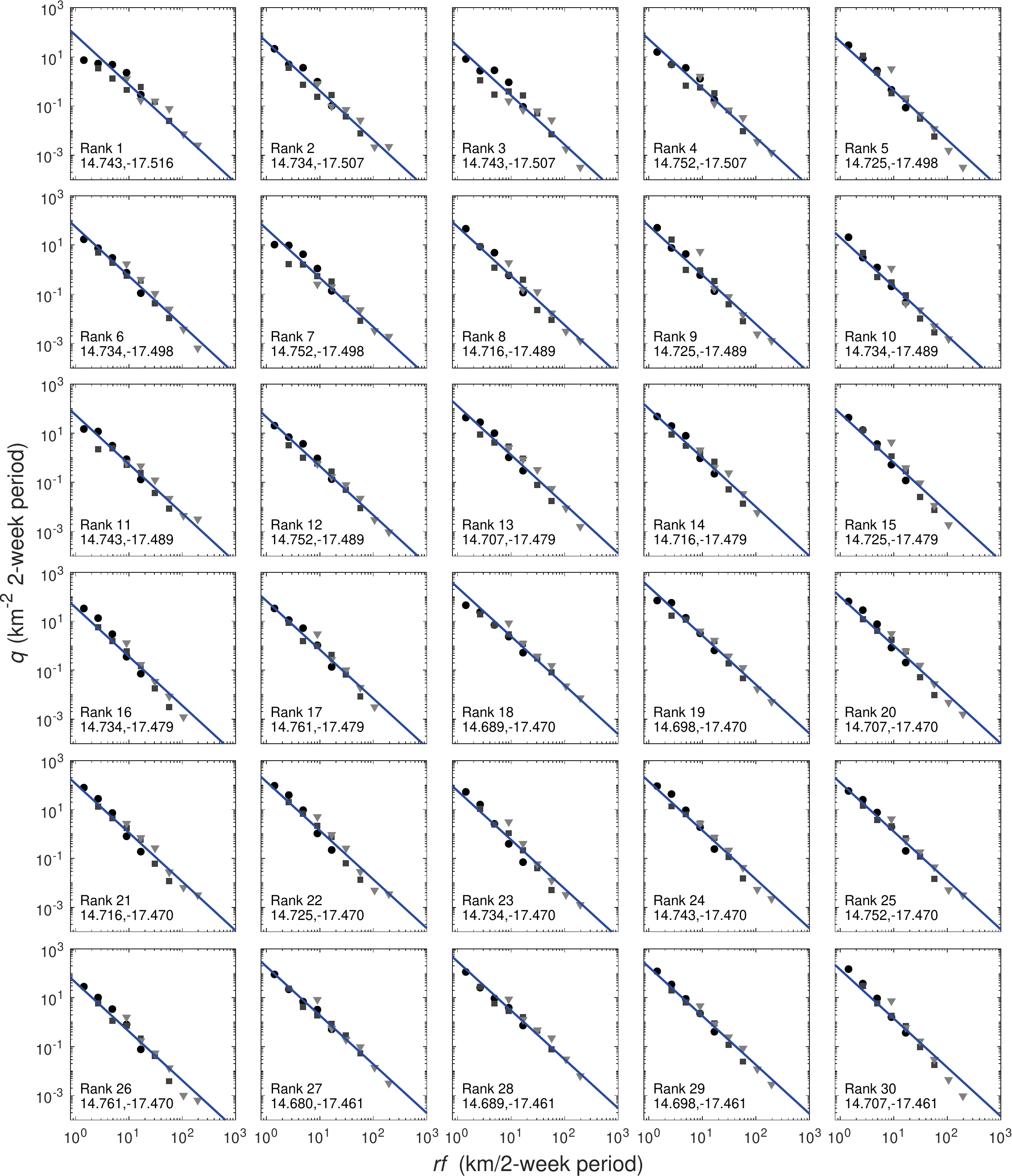}\vspace{-0.2cm}
\caption{\label{fig:indScalingSenI}{\footnotesize{\bf Universality of the scaling relation \boldmath${q \propto (rf)^{-2}}$ across Senegal.} The panels depict the data for individual locations (1$\,$km $\times$ 1$\,$km grid cells), ranked according to the total number of visitors from neighbouring cells. Shown are locations of rank 1-30 (from top left to bottom right). Symbols denote different frequency bins, with \protect\markerone~for $[1,2)$, \protect\markertwob~for $[2,5)$ and \protect\markerthreeb~for $[5,10)$ visits per 2-week period. The geographic coordinates of each location (latitude and longitude of the centre point of the grid cell) are indicated. The straight lines denote the theoretically predicted inverse square of $rf$.}}
\end{figure}

\clearpage
\thispagestyle{empty}
\begin{figure}
\centering
\includegraphics[width=0.95\textwidth]{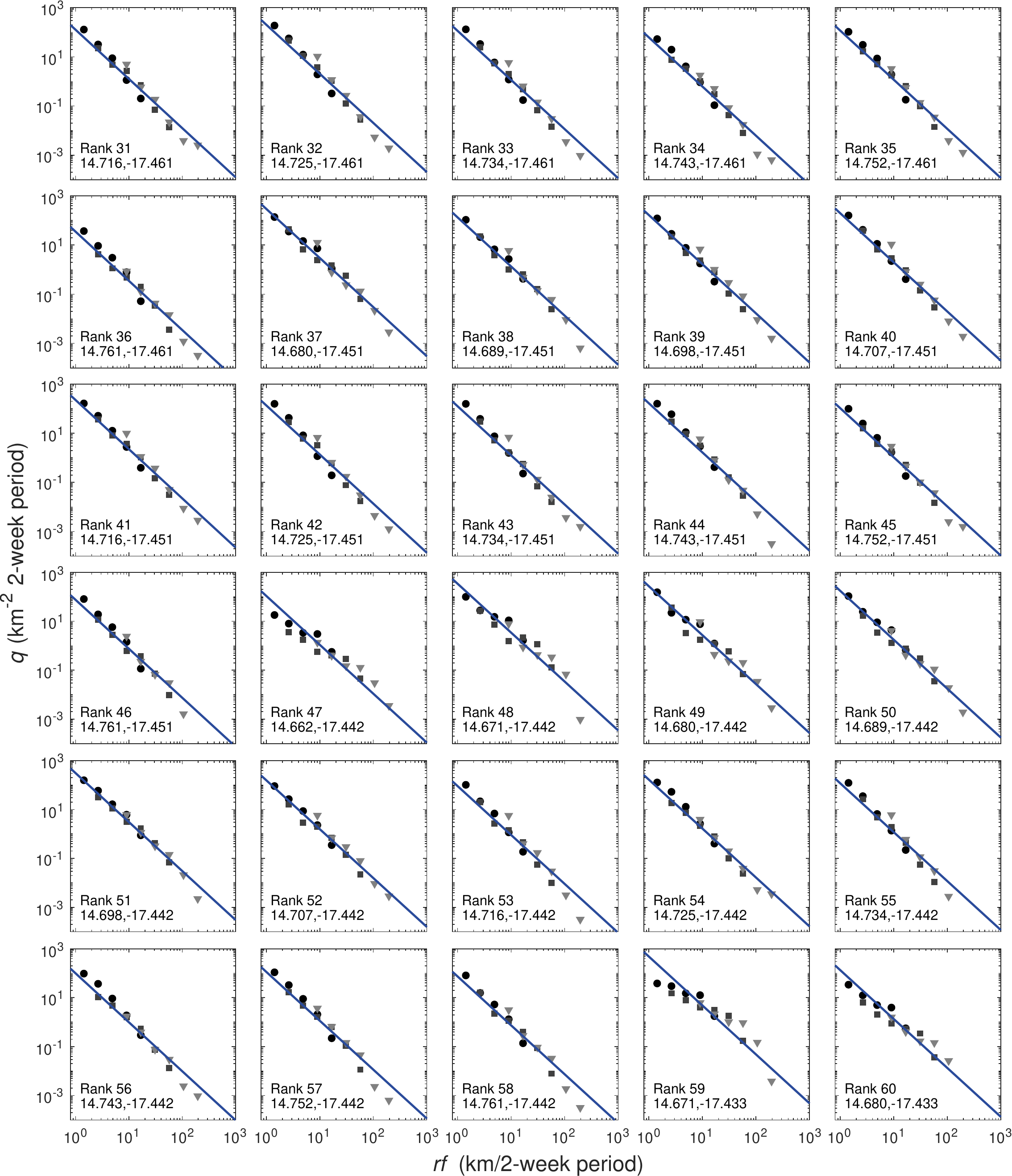}\vspace{-0.2cm}
\caption{\label{fig:indScalingSenII}{\footnotesize{\bf Universality of the scaling relation \boldmath${q \propto (rf)^{-2}}$ across Senegal.} The panels depict the data for individual locations (1$\,$km $\times$ 1$\,$km grid cells), ranked according to the total number of visitors from neighbouring cells. Shown are locations of rank 31-60 (from top left to bottom right). Symbols and notations are as in \mbox{Fig.~\ref{fig:indScalingSenI}}.}}
\end{figure}

\clearpage
\thispagestyle{empty}
\begin{figure}
\centering
\includegraphics[width=0.95\textwidth]{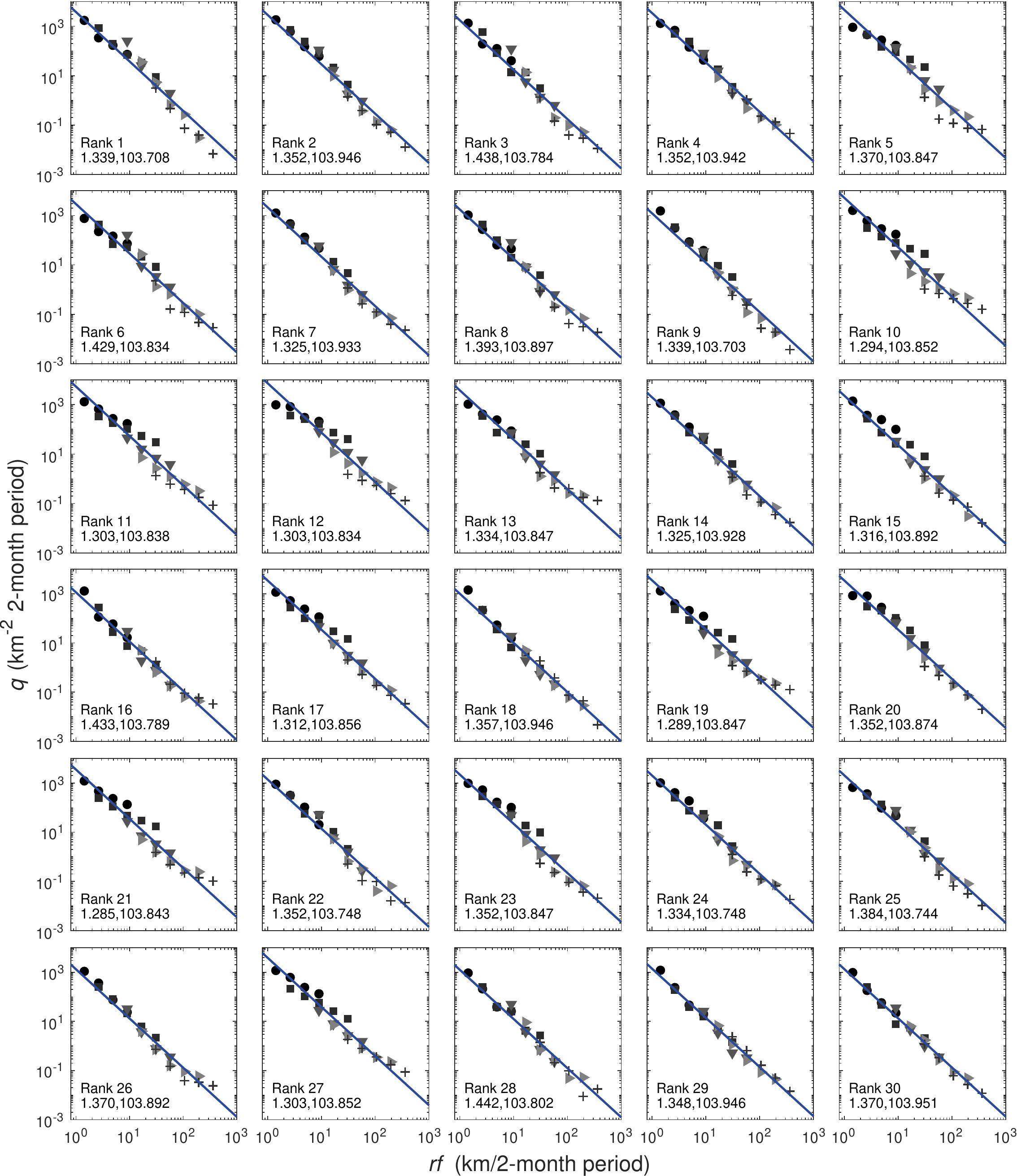}\vspace{-0.2cm}
\caption{\label{fig:indScalingSingI}{\footnotesize{\bf Universality of the scaling relation \boldmath${q \propto (rf)^{-2}}$ across Singapore.} The panels depict the data for individual locations (500$\,$m $\times$ 500$\,$m grid cells), ranked according to the total number of visitors from neighbouring cells. Shown are locations of rank 1-30 (from top left to bottom right). Symbols denote different frequency bins, with \protect\markerone~for $[1,2)$, \protect\markertwo~for $[2,5)$, \protect\markerthreec~for $[5,10)$, \protect\markersixb~for $[10,20)$ and $\bm{+}$~for $[20,40)$ visits per 3-month period. The geographic coordinates of each location (latitude and longitude of the centre point of the grid cell) are indicated. The straight lines denote the theoretically predicted inverse square of $rf$.}}
\end{figure}


\clearpage
\thispagestyle{empty}
\begin{figure}
\centering\hspace{-1cm}
\includegraphics[width=0.75\textwidth]{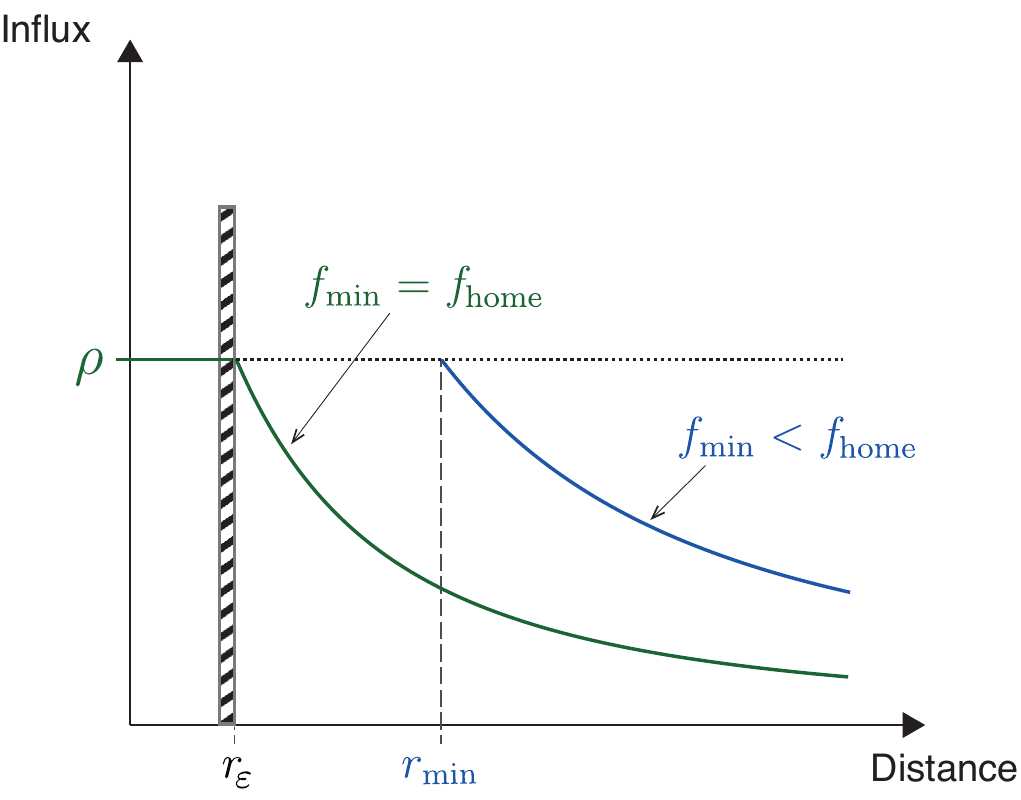}
\caption{\label{fig:model_boundary_condition}{\footnotesize{\bf Estimation of the magnitude of flows from population density \boldmath$\rho$.} The schematics shows a zoom-in on the immediate vicinity of an attracting location (small values of $r$), where it is reasonable to assume that $\rho \approx \text{const}$. Hence, the local population density imposes an upper bound on the influx, \mbox{$\int \! q \ud f \leq \rho$}. A simple boundary condition of our continuous model then dictates that the minimum visiting frequency of all individuals living directly on the boundary (each being assigned to a point at $r=r_\varepsilon$) assumes the minimum frequency with which all individuals living inside the attracting location return home, $f_{\text{min}} = f_{\text{home}} \approx 1\,\text{day}^{-1}$. The minimum distance for locations from which individuals visit with minimum frequency $f_{\mathrm{min}} < f_{\text{home}}$ is $r_{\text{min}} = r_\varepsilon \sqrt{ f_{\text{home}} / f_{\text{min}}}$.}}
\end{figure}
\end{document}